\documentclass[12pt]{article}
\global\parskip 6pt
\newcommand{\be}{\begin{eqnarray}}
\newcommand{\ee}{\end{eqnarray}}

\newcommand{\lb}{\label}
\def\>{\rangle}
\def\<{\langle}

\begin{document}
\begin{titlepage}
\hfill{hep-th/0212308}
\vspace*{.5cm}
\begin{center}
{\large{{\bf Relaxation in Conformal Field Theory, Hawking-Page Transition,\\[.5ex]
and Quasinormal/Normal Modes}}} \\
\vspace*{1.5cm}
Danny Birmingham\footnote{Email: birm@itp.stanford.edu; On leave from:
Department of Mathematical Physics, University College Dublin,
Ireland}\\
\vspace{.1cm}
{\em Department of Physics,
Stanford University\\
Stanford, CA 94305-4060\\
USA}\\
\vspace*{.5cm}
Ivo Sachs\footnote{Email: ivo@maths.tcd.ie}\\
\vspace*{.1cm}
{\em School of Mathematics,
Trinity College Dublin\\
Dublin 2\\
Ireland}\\
\vspace*{.5cm}
Sergey N. Solodukhin\footnote{Email:
soloduk@theorie.physik.uni-muenchen.de}\\
\vspace*{.1cm}
{\em Theoretische Physik,
Ludwig-Maximilians Universit\"{a}t\\
Theresienstrasse 37\\
D-80333, M\"{u}nchen\\
Germany}\\
\vspace*{1cm}
\begin{abstract}
\noindent {We study the process of relaxation
back to thermal equilibrium in $(1+1)$-dimensional conformal
field theory at finite
temperature.  When the size of the system is much larger than the
inverse temperature, perturbations decay exponentially with time.
On the other hand, when the inverse temperature is large, the
relaxation is oscillatory with characteristic period set by the
size of the system. We then analyse the intermediate regime in two
specific models, namely free fermions, and a strongly coupled
large ${\tt k}$ conformal field theory which is dual to string theory on
$(2+1)$-dimensional anti-de Sitter spacetime. In the latter case,
there is a sharp transition between the two regimes in the 
${\tt k}=\infty$ limit,
which is a
manifestation of the gravitational Hawking-Page phase transition.
In particular,  we establish a direct connection between
quasinormal and normal
modes of the gravity system, and the decaying and oscillating
behaviour
of the conformal field theory.   \\
\vspace*{.25cm}
}
\end{abstract}
\vspace*{.25cm}
December 2002
\end{center}
\end{titlepage}

\section{Introduction} An important problem in finite temperature field
theory and statistical mechanics is to study the response of
a system in  thermal equilibrium to a generic perturbation. In
particular,
one is typically interested in
understanding the process of relaxation back to thermal equilibrium.
For small
perturbations, this is well described by linear response
theory, and the time evolution of the relaxation is determined by
the retarded correlation function of the perturbation
\cite{Fetter,lebellac,kapusta}. Generally, in the presence of
interactions
this problem is addressed within finite temperature perturbation
theory.
A special role is played by scale invariant theories, where
the zero temperature $2$-point functions are uniquely determined
(up to a normalisation) by scale invariance. However, finite
temperature
introduces a new scale and consequently the conformal Ward identities
no longer determine the Green's functions completely.
However, significant progress can be made if
the conformal field theory (CFT) has a dual formulation in terms of
gravity
(or string theory) on anti-de Sitter space (AdS) spacetime.
Indeed, the AdS/CFT
\cite{Maldacena:1997re,Gubser:1998bc,Witten:1998qj,Aharony:1999ti}
duality predicts that the retarded CFT correlation functions are in
1-1 correspondence with Green's functions on anti-de Sitter space with
appropriate boundary conditions
\cite{Danielsson:1999zt,KalyanaRama:1999zj,Horowitz:1999jd,Birmingham:2001pj,Son:2002sd}.
Furthermore,
the poles of the retarded CFT correlators are given by the quasinormal
modes in AdS \cite{Danielsson:1999zt,Birmingham:2001pj}.
This correspondence was verified
explicitly in the high temperature regime of the 2-dimensional CFT,
dual to supergravity on AdS$_3$ \cite{Birmingham:2001pj}.
Other calculations of quasinormal modes in anti-de Sitter backgrounds
have
appeared in \cite{QNM}.

The purpose of the present paper is to analyse finite volume effects
for a quantum conformal field theory living
in one space-like dimension (closed to form a circle of length $L$).
In two limiting cases, when
the dimensionless parameter $LT$ is infinite or zero, conformal
invariance
completely determines the correlation function, independent of the
details of the theory in question. The behaviour of the retarded
correlation functions in these two cases is qualitatively different.
In the first case, the perturbation decays exponentially with
characteristic
time proportional to the inverse temperature.  In the second case,  we
have oscillatory behaviour with characteristic period determined
by $L$. Our main purpose here is to study the relaxation process in the
intermediate regime when $LT$ changes from zero to infinity.

After
describing the qualitative behaviour based on general arguments,
we analyse quantitatively the case of a non-interacting theory, and
contrast its behaviour
with  the strongly coupled large ${\tt k}$ CFT dual to supergravity on
AdS$_3$ (see also \cite{Maldacena:2001kr,Maldacena:1998bw}).
(The parameter ${\tt k}$ plays the role of $N$ in the usual terminology of 
large $N$ CFT dual to AdS gravity.)
In the latter case,  we have an explicit expression for the 2-point
function
at finite temperature and finite volume.
We can thus analyse  the
linear response of the CFT.  In the limit ${\tt k} = \infty$, there is a sharp transition
between a regime of exponential decay and oscillation. This is a
manifestation
of the gravitational Hawking-Page phase transition between the BTZ black hole
and thermal AdS.
Moreover, we then establish a direct connection between the behaviour
of
the linear response,  and the behaviour of the corresponding bulk
perturbations.
In particular, we show that the regime of exponential decay is governed
by the
quasinormal modes of the BTZ black hole, while the regime of
oscillation
is governed by the normal modes of thermal AdS.
Thus, the behaviour
of the bulk AdS perturbation is mirrored precisely by the behaviour of
the linear response
of the CFT.
Finally, we speculate on 
the expected behaviour of the strongly coupled CFT 
at finite values of $\tt k$.

\section{Linear Response Theory}
We consider a system initially in thermal equilibrium, and apply
a small perturbation.
The main goal of linear response theory is to study the
change in the expectation value
of an operator ${\cal O}(x,t)$  as a result of this perturbation.
The total Hamiltonian of the system takes the form
$H^{\prime}(t) = H + H_{{\rm ext}}(t)$,
where $H$ is the unperturbed Hamiltonian, and $H_{{\rm ext}}(t)$
couples an external field
to the system, with the assumption that
$H_{{\rm ext}} = 0$, for $t < t_{0}$.
The change in the ensemble average of ${\cal O}$ is given by
\be
\delta\langle{\cal O}(x,t)\rangle = i \int_{t_{0}}^{t} dt^{\prime}\;
{\rm{Tr}}\{\hat{\rho}\;[H_{{\rm ext}}(t^{\prime}), {\cal O}(x,t)]\}\ ,
\lb{1}
\ee
where $\hat{\rho}$ is the unperturbed  thermal density matrix.
If we take the perturbation to be
\be
H_{{\rm ext}}(t) = \int dx\;J(x,t){\cal O}(x,t)\ ,
\ee
where $J(x,t)$ is the external source, then (\ref{1}) takes the form
\be
\delta\langle{\cal O}(x,t)\rangle = \int_{-\infty}^{\infty}dt^{\prime}
\int dx^{\prime}\;J(x^{\prime}, t^{\prime})D^{R}(x,t;x^{\prime},
t^{\prime})\ ,
\ee
where
\be
D^{R}(x,t;x^{\prime},t^{\prime}) = -i\theta(t-t^{\prime}){\rm
Tr}\{\hat{\rho}
\;[{\cal O}(x,t), {\cal O}(x^{\prime}, t^{\prime})]\}
\ee
is the retarded propagator.

For the particular case of an instantaneous perturbation of the form
\be
J(x,t) = \delta(t) e^{ikx}\ ,
\ee
one finds
\be
\delta \langle {\cal O}(x,t)\rangle = e^{ikx} \int_{-\infty}^{\infty}
\frac{d\omega}{2 \pi} e^{-i\omega t}D^{R}(\omega, k)\ .
\lb{6}
\ee
In the following, it will be useful to recall the Lehmann
representation of the retarded propagator
\be
D^{R}(\omega, k) = \int_{-\infty}^{\infty}\frac{d \omega^{\prime}}{2
\pi}
\frac{\rho(\omega^{\prime}, k)}{\omega - \omega^{\prime} + i \epsilon}\,
\ee
where the spectral function
$\rho(\omega,k)$ is the Fourier transform of the commutator
$\rho(x,t;x^{\prime},t^{\prime})
= \langle [{\cal O}(x,t),{\cal O}(x^{\prime}, t^{\prime})]\rangle$.

{}From general arguments,
we know that $D^R(\omega,k)$ is an
analytic function in the upper half
$\omega$-plane.
If the energy levels of the system are discrete and the ensemble sum
converges,
then we conclude that $D^{R}(\omega, k)$ is a meromorphic function
of $\omega$ in the lower half plane,  with simple poles on the
real axis. Correspondingly, as can be seen from (\ref{6}),
the linear response will show oscillatory
behaviour. Indeed, as shown in \cite{Dyson:2002nt}-\cite{Kleban2}, the energy spectrum
for a system with finite entropy is discrete, and will in general exhibit the phenomenon
of Poincar\'{e} recurrence.  If the
energy levels are continuous, the properties of $D^{R}(\omega, k)$
can be more complicated. At zero temperature, the retarded Green's
function will generally have poles and cuts on the real axis,
corresponding
to stable states and multi-particle states, respectively. Furthermore,
there
can also be poles in the lower half $\omega$-plane corresponding to
resonances.
The distance of the poles from the real line then determines the decay
time of
such a resonance. At finite temperature, $D^{R}(\omega, k)$ is a
meromorphic function in the lower half plane. Eqn.
(\ref{6}) then shows that the characteristic times for the
thermalisation of an instantaneous
perturbation is determined by the imaginary part
of the poles of $D^{R}(\omega, k)$ in the complex $\omega$-plane.
For a generic interacting theory,  the location of these poles will
depend
non-trivially on the coupling constants,  and concrete results are
known only
in specific limits, where perturbation theory is applicable.

\section{Conformal Field Theory}
To see how these general results are realized in $(1+1)$-dimensional
conformal
field theory, we first
calculate the correlation functions on a torus
with periods $L$ and $\beta=T^{-1}$. The correlation function in real
time,
$t$, is
then obtained by the analytic continuation $\tau=it$,
where $\tau$ is imaginary time.   When the size of one
direction on the torus is taken to infinity, the torus becomes a
cylinder.
Using Cardy's result \cite{Cardy:ie}, the correlation function on the
cylinder
is  obtained from the  correlator on the plane by the conformal
mapping\footnote{Note that in $2$-dimensional CFT the left and right
sectors are
independent. In particular, they can be defined on cylinders with
different radii.}
$w= (l/2\pi )\ln z$, $\bar w = (\bar l/2\pi )\ln \bar z $,
where $w$ and $z$ are complex coordinates on
the cylinder and the plane, respectively. The $2$-point function on the
cylinder is then given by  \cite{Cardy:ie}
\be
\<{\cal{O}}(w,\bar{w}){\cal{O}} (w',\bar{w}') \>
={(\pi/l)^{2h} (\pi/\bar l)^{2\bar{h}}\over
[\sinh (\pi/l) (w-w')]^{2h}[\sinh
(\pi/\bar l) (\bar{w}-\bar{w}')]^{2\bar{h}}}\ ,\lb{2}
\ee
where $(h,\bar{h})$ are the conformal weights of ${\cal{O}}$.

Suppose now that the temperature is finite, while the size $L$ is
taken to infinity; then $l,\bar l=\beta(1\pm\mu)$, where $\mu$ is the
chemical potential and $w=\sigma+i \tau$. After
analytic continuation $\tau=it$, we have $w=\sigma-t$ and
$\bar{w}=\sigma+t$. We then obtain the real time correlation
function
\be
D_\pm(t,\sigma) &=&
\<{\cal{O}}(t\mp i\epsilon,\sigma ){\cal{O}}(0,0 )\> \nonumber \\
&=&{(\pi T_R)^{2h} (\pi T_L)^{2\bar{h}}\over [\sinh \pi T_R
(\sigma-t\pm i\epsilon)]^{2h}
[\sinh \pi T_L (\sigma+t\mp i\epsilon   )]^{2\bar{h}}}\ , \lb{3}
\ee
where $T_{R}\!=\! 1/l$ and $T_{L}\!=\!1/\bar{l}$
are the effective right and left temperatures.
Clearly, the correlation function decays exponentially for large time
$t$;
for example, $D_\pm(t,0)\sim
e^{-2\pi T(h+\bar{h}) t}$, when $T_{L} = T_{R} = T$.

For $\{h,\bar h\}\in
\frac{1}{2} {\bf N}$, the Fourier transform of the commutator,
$\rho=(D_+- D_-)$, can be evaluated by the method of
residues leading to
\begin{equation}\label{FT1}
\rho(\omega, k)\propto
\left|\Gamma\left(h+i\frac{p_-}{2\pi T_R}\right)
\Gamma\left(\bar h+i\frac{p_+}{2\pi T_L}\right)\right|^2 \ ,
\end{equation}
where $p_\pm=\frac{1}{2}(\omega\mp k)$. From (\ref{FT1}),
we see that
$\rho(\omega, k)$ has an infinite set of simple poles on either
side of the real line.
The poles of the Green's function $D^R(\omega,k)$ are then
obtained by
restricting this set to the lower half plane.
These are given by \cite{Birmingham:2001pj}
\begin{eqnarray}\label{PFD1}
\omega&=&k-4\pi i\; T_L(n+\bar h)\ ,\nonumber\\
\omega&=&-k-4\pi i\; T_R(n+h)\ , \;\;n\in{\bf N}\ .
\lb{qnm}
\end{eqnarray}

On the other hand, if the size $L$ is kept finite and
the inverse temperature $\beta=T^{-1}$ is taken
to infinity  then $l=L$. The direction of the imaginary axis can
be chosen in the $\sigma$ direction, $w=\tau +i\sigma$. The
analytic continuation to real time now results in
$w=i(t+\sigma )$ and $\bar{w}=i(t-\sigma )$. Thus, we find that
(\ref{2})
leads to the real time correlation function
\be
D_\pm(t,\sigma )
&=&\<{\cal{O}}(t\mp i\epsilon,\sigma ){\cal{O}} (0,0 )\> \nonumber \\
&=&{(\pi/iL)^{2(h+\bar{h})} \over
[(\sin \frac{\pi}{L}  (t \mp i \epsilon+\sigma )]^{2h}[(\sin
\frac{\pi}{L}
(t \mp i \epsilon-\sigma )]^{2\bar{h}}}\ .
\lb{4}
\ee
This clearly shows the expected oscillatory behaviour in real time.
In the following, we
specialise to the subset of operators with $h=\bar h\in {\bf N}/2$. To
isolate the pole structure of the retarded Green's function, we again
consider
the Fourier transform of the
commutator
\be
\rho(\omega, k) = \int_{-\infty}^{\infty} dt \int_{0}^{L}d \sigma\;
e^{i\omega t - ik\sigma} (D_{+} - D_{-})\ ,
\ee
with $k=\frac{2\pi}{L}m$.
We first consider the case with $h = \bar{h} = 1/2$. Then $\rho$ takes
the simple form
\be
\rho(\omega, k)
\propto \sum_{n \in {\bf Z}}\delta \left(\frac{\omega L}{4 \pi} -
\frac{|m|}{2}
-h - n\right)\left[\theta\left(\frac{\omega L}{2 \pi} -|m|\right) -
\theta \left(-\frac{\omega L}{2 \pi} -|m|\right)\right]\ .
\lb{spec}
\ee
Thus, we find the poles
\be\label{set}
\omega &=& \frac{2 \pi}{L}|m|  + \frac{4 \pi}{L}(n+h)\ ,\nonumber \\
\omega &=& -\frac{2\pi}{L}|m| - \frac{4 \pi}{L}( n + h)\ ,
\lb{poles}
\ee
with $m \in {\bf Z}$, and $n \in {\bf N}$.
The spectral function for higher values of $h$ takes a form similar
to (\ref{spec}), with additional
prefactors designed such that the pole
structure is given by (\ref{poles}).

The considerations so far are quite general, and rely only on the
conformal properties of the operators driving the perturbation.
The behaviour of the correlation functions (\ref{3}) and (\ref{4})
in the two limiting cases is thus universal, and in agreement with
general expectations.
However, the
behaviour in the intermediate regime, when both $L$ and $T$ are finite,
cannot be derived entirely from
conformal symmetry.
In order to understand this regime, we first consider the $2$-point
function of free fermions on the torus \cite{DiFrancesco:nk}.
This takes the form,
\be
\<\psi (w)\psi(0)\>_{\nu}={\theta_\nu \left(\frac{w}{\beta}\mid i
\frac{L}{\beta}\right)\partial_{w}
\theta_{1}\left(0\mid i\frac{L}{\beta}\right)\over
\theta_{\nu}\left(0\mid i\frac{L}{\beta}\right)\theta_1
\left(\frac{w}{\beta} \mid i \frac{L}{\beta}\right)}\ ,
\lb{5}
\ee
where $w=\tau+i\sigma$ and $\nu$ characterises the
boundary conditions for $\psi (w)$. For finite temperature boundary
conditions, we have $\nu=3,4$. Using the properties of
$\theta$-functions, it is then easy to see that (\ref{5})
is invariant under shifts $w\rightarrow w+\beta$
and $w\rightarrow w+iL$.

The real time correlation function
is obtained from (\ref{5}) by the substitution $w=i(t+\sigma )$.
The resulting real time correlator
is thus a periodic function of $t$ with period $L$. Zeros of the
theta function $\theta_1(w/\beta|iL/\beta)$ are located at
\cite{DiFrancesco:nk}  $w=m\beta +inL$, where $m,n$ are
integers. Therefore, for real time $t$, the correlation
function (\ref{5}) is a sequence of singular peaks located at
$(t+\sigma)=nL$. In the limit $L/\beta\rightarrow \infty$, the
correlation function
(\ref{5}) approaches
(\ref{3}) (actually the left-moving part of (\ref{3}) with $h=1/2$)
which exponentially decays with time.
In the opposite limit, when $L/\beta\rightarrow 0$,
it approaches the oscillating function (\ref{4}). This is in agreement
with our discussion using Cardy's arguments. It is important to observe
that
the exponential decay
of the function in the limit of large $L$ is  consistent with
periodicity of the real time correlation function (\ref{5})
with period $L$. In order to see this, we note that the correction to
the
leading behaviour is governed by $e^{-{\pi
L/\beta}}e^{\pi(t+\sigma)/\beta }$.
Thus, the limit of large $L$ is meaningful only for times much smaller
than $L$.
For such $t$, the exponentially decaying function is a good description
of the correlation function. However, for $t$ approaching $L$ the
correction terms become important, and the
time-periodicity of the correlation function (\ref{5}) is restored.

\section{Strong Coupling Regime} At strong coupling, conformal symmetry
is not enough to
determine the linear response at finite volume. We are therefore unable
to obtain exact
results in the general case. However,  progress can be made in some
special cases.
Here, we consider the supersymmetric conformal field theory dual to
string
theory on AdS$_3$. This theory describes the low energy excitations of
a large number of D1- and D5-branes
\cite{Maldacena:1997re,Aharony:1999ti,Maldacena:1998bw}.
It can be interpreted as a gas of
strings that wind around a circle of length $L$ with target space
$T^4$.
The individual strings can be simply or multiply wound such that the
total
winding number is ${\tt k}=\frac{c}{6}$, where $c>\!>1$ is the central
charge. The parameter ${\tt k}$ plays the role of $N$ in
the usual terminology of large $N$ CFT dual to AdS gravity.
In order to obtain information
about the correlation functions in the strong coupling regime of this
theory,
we can appeal to the AdS/CFT correspondence.
According to this duality,
each supergravity perturbation $\Phi_{(m,s)}$ of mass $m$ and
spin $s$ propagating on AdS$_3$ has a corresponding
operator ${\cal{O}}_{(h,\bar{h})}$ in the dual conformal field theory.
This operator is characterised by conformal weights $(h,\bar{h})$, with
$h+\bar{h}=\Delta$, $h-\bar{h}=s$, and $\Delta$ is determined in
terms of the mass of the field. The CFT correlators are then determined
in
terms of the corresponding bulk Green's functions.
In addition, there is a correspondence between (quasi)normal modes in
the gravitational
background and the poles of the retarded Green's functions in the
conformal field
theory \cite{Danielsson:1999zt}.

According to the original prescription \cite{Witten:1998qj}, each AdS
space
which asymptotically approaches the given
two-dimensional manifold should contribute
to the calculation, and one thus has to sum over all such spaces.
In the case of interest, the two-manifold is a torus $(\tau, \sigma )$,
where $\beta$ and $L$ are the respective periods.
There is an $SL(2, {\bf Z})$ family of AdS$_3$ spaces which approach
the torus asymptotically
\cite{Maldacena:1998bw,Dijk}.
For the purpose of understanding the Hawking-Page phase transition, it suffices to consider
the BTZ black hole  and thermal AdS space,
corresponding to anti-de Sitter space filled with thermal radiation.
Both spaces can be represented \cite{Carlip:1994gc}
as a quotient of three-dimensional
hyperbolic space $H^3$, with line element
\be
ds^2={1\over y^2}\;(dzd\bar{z}+dy^2)\ ,\ y>0\ .
\lb{7}
\ee
The BTZ black hole
(for simplicity we consider only the non-rotating BTZ black hole) has
inverse temperature
$\beta=2\pi/r_{+}$, where $r_+$ is the horizon radius.
Also,
$z \sim e^{2\pi w/\beta}$ and $y \sim e^{2 \pi \sigma/\beta}$, where
$w=\sigma+i\tau$.
Thus, the orbifold identification is given by
\be
z \rightarrow e^{2\pi L/\beta}z\ ,\;\; y \rightarrow e^{2 \pi
L/\beta}y\ .
\ee
For the thermal AdS at the same temperature $\beta^{-1}$, we have
$z \sim e^{2\pi w/L}$ and $y \sim e^{2 \pi \tau/L}$, where
$w=\tau+i\sigma$. In this case, the identification is
\be
z \rightarrow e^{2\pi \beta /L}z\ , \;\;y \rightarrow e^{2 \pi
\beta/L}y\ .
\ee
In both cases, the boundary of
the three-dimensional space is a torus with
periods $L$ and $\beta$. In particular, the two configurations
(thermal AdS and the BTZ black hole) are T-dual to
each other, and are obtained by the interchange of the coordinates
$\tau \leftrightarrow \sigma$ on the torus.

The correlation function of the dual operators then contains the sum of
two contributions \cite{Maldacena:2001kr} as
\be
\<{\cal{O}}(w,\bar{w}) {\cal{O}} (w',\bar{w}')\>_{\rm Torus}
=
e^{-S_{\rm BTZ}}\<{\cal{O}} ~{\cal{O}}'\>_{\rm BTZ}+
e^{-S_{\rm AdS}}\<{\cal{O}} ~{\cal{O}}'\>_{\rm AdS}\ , \lb{8}
\ee
where
\be
S_{\rm BTZ}=-{\tt k} \pi \frac{L}{\beta}\ ,
\;\;S_{\rm AdS}=-{\tt k}  \pi \frac{\beta}{L}\ ,
\ee
are the Euclidean actions of the
BTZ black hole and thermal AdS$_3$, respectively
\cite{Maldacena:1998bw}.
We see that these contributions are dual to each other under
interchange of  $\beta$ and $L$. Using the AdS/CFT correspondence, the
correlation functions on the torus were computed in
\cite{Keski-Vakkuri:1998nw,Corrfunct} for
the case of a scalar field ($h=\bar h$).
For the BTZ background, the result takes the form
\cite{Keski-Vakkuri:1998nw}
\be
\< {\cal{O}}(w,\bar{w}) {\cal{O}} (0,0)\>_{\rm BTZ}
= \sum_{n\in {\bf Z}}{1\over \sinh[\frac{\pi}{\beta}(w+nL)]^{2h}
\sinh[\frac{\pi}{\beta}(\bar{w}+nL)]^{2h}}\ ,
\lb{9}
\ee
where $w=\sigma+i\tau$.
Note that (\ref{9}) takes the form of the strip expression (\ref{2})
summed over images to make it doubly periodic. On the supergravity
side, the
justification to sum over images comes simply from the fact that the
correlator solves a Green's function equation.  From the CFT point of
view,
this result is non-trivial. Indeed,  for a generic
CFT, summing over images does not produce
the correct finite volume correlator.
For example, the free field correlator
(\ref{5}) does not have this form.

Expression (\ref{8}) is the result for the correlation
function in the strong coupling regime.
Although each term in the sum (\ref{8}) is not modular invariant,
the sum over the full $SL(2, {\bf Z})$ family does have this property
\cite{Maldacena:1998bw,Dijk}.
Depending on the ratio
$L/\beta$, one of the two terms in (\ref{8}) dominates
\cite{Maldacena:1998bw}.
For high temperature ($L/\beta$ is large)
the BTZ is dominating, while at low temperature  ($L/\beta$ is small)
the thermal AdS is dominant. The transition between the two regimes
occurs at $\beta=L$. In terms of the gravitational physics, this
corresponds to the
Hawking-Page phase transition \cite{Hawking:1982dh}.
This is a sharp transition in the limit ${\tt k} = \infty$,
which is the case when the supergravity description is valid.
In this limit, the BTZ black hole is the sole contribution
for $L>\beta$, while thermal AdS is the only term which contributes
for $L < \beta$.

The two terms in (\ref{8}) have drastically different behaviour
as functions of real time. After the analytic continuation
$\tau=it$, the BTZ contribution (\ref{9}) produces the correlator
\be
\<{\cal{O}} (t,\sigma ){\cal{O}} (0,0)\>_{\rm BTZ}
=\sum_{n\in {\bf Z}}{1\over
[\sinh \frac{\pi}{\beta} (\sigma-t+nL )]^{2h}
[\sinh \frac{\pi}{\beta} (t+\sigma+nL )]^{2h}}\ ,\lb{10}
\ee
which exponentially decays with time. Furthermore,
in the limit $L/\beta\rightarrow \infty$, only the $n=0$ term
in (\ref{10}) contributes,  and we thus recover the universal behaviour
(\ref{3}).

It is clear that the Fourier transform of the spectral function for
(\ref{10}) is again given by
(\ref{FT1}), with $k$ restricted to the discrete values
$\frac{2\pi}{L}m$.
Thus, the poles of $D^{R}$ are given by (\ref{qnm})
with $k = \frac{2 \pi}{L}m$. At first sight, this appears to be in
contradiction with the general behaviour discussed in section 2.
Indeed, for finite volume we
expect a discrete energy spectrum and consequently that the poles of
$D^R$ should lie on the real axis. The resolution of this puzzle lies
in the peculiar properties of the CFT under consideration
\cite{Aharony:1999ti}. For $L/\beta>1$, the
typical configuration consists of a relatively small number of multiply
wound
strings so that the effective volume relevant for the energy gap is
$L_{\rm eff}\simeq {\tt k} L\to\infty$.
Of course, this explanation immediately raises
another puzzle: If the effective volume of the theory is infinite, how
come
that the correlation function is periodic in $\sigma$ with period $L$.
The reason for this lies in the fact that the
operator ${\cal{O}}$ does not distinguish between
simply wound and multiply wound strings. Consequently,  ${\cal{O}}$
still
sees a finite volume.

For $h\neq \bar h$, which corresponds to fields of non-zero spin in
AdS$_3$,
the finite temperature Green's functions have not been worked out in
the
literature. However, it was shown in \cite{Birmingham:2001pj}
that there is a 1-1 correspondence between the poles of the retarded
finite
temperature Green's function and the quasinormal modes for fields of
spin $s=h-\bar h$ in the BTZ background. Quasi-normal modes are
solutions
to the wave equations which are purely ingoing at the horizon and
subject to the boundary condition that the current  vanishes at
infinity\footnote{Originally,  the quasinormal modes were required to
satisfy Dirichlet conditions at asymptotic infinity
\cite{Horowitz:1999jd}.
However, as shown in \cite{Birmingham:2001pj},
this leads to problems for $h\le 1$ which can be resolved by
requiring the vanishing of the current.}.
Only a discrete set of modes satisfying  these boundary
conditions is possible,  and the frequencies are
shown in \cite{Birmingham:2001pj} to be identical with (\ref{qnm}).
In \cite{Birmingham:2001pj},
the correspondence was shown for $L/\beta=\infty$. However, the above
discussion shows that this is in fact valid for $L/\beta>1$.
Indeed, one finds that the bulk perturbation decays via these
quasinormal modes
in precisely the same way as the linear response of the conformal field
theory
given by (\ref{6}).

The question
then arises as to the behaviour for $L/\beta<1$. On the gravity side,
(\ref{8}) implies that the thermal AdS Green's function gives the
relevant
contribution to the real time correlation function.
The result for the thermal AdS is obtained from (\ref{9}) under the
interchange $\tau \leftrightarrow \sigma$ and $\beta \leftrightarrow
L$.
Hence, we have \cite{Corrfunct}
\be
\<{\cal{O}} (t,\sigma ){\cal{O}} (0,0)\>_{\rm AdS}
=\sum_{n\in {\bf Z}}{1\over
[\sin {\pi\over L}  (t+\sigma +i\beta n)]^{2h}
[\sin {\pi\over L}  (t-\sigma +i\beta n)]^{2h}}  ~~,
\lb{11}
\ee
which is periodic in $t$ with period $L$. Clearly, it represents  a
periodic
sequence of singular peaks at $t\pm\sigma =nL$.
In the limit of infinite
$\beta/L$, only the $n=0$ term contributes and (\ref{11})
approaches the expression (\ref{4}).
The Hawking-Page transition is thus a transition between
oscillatory behaviour at low temperature  and  exponentially decaying
behaviour at high temperature. From the CFT point of view, this
behaviour is
explained by the fact that in the low temperature phase the
generic configuration is given by simply wound strings so that the
effective volume is finite and consequently the energy spectrum is
discrete
in agreement with general argument.

For $h\in\frac{1}{2}{\bf N}$, the Fourier transform of the spectral
function
for the finite
temperature AdS Green's function
(\ref{11}) is again given by (\ref{spec}), with
simple poles given by (\ref{poles}).
On the AdS side, the set of frequencies (\ref{poles}) is
identical to the set of normalisable  modes
\cite{Balasubramanian:1998sn}: these modes are regular at the origin
of AdS$_3$ and are normalisable at infinity. As a result, the
oscillating behaviour
of the bulk perturbation is mirrored by the oscillating behaviour of
the linear response.
In \cite{Balasubramanian:1998sn},
the normalisable modes were obtained for a scalar field $h=\bar h$.
However, as shown in
\cite{Metsaev:1999ui},
higher spin $s$ equations of motion in AdS$_3$ can be reduced
to that of massive scalar fields for any $s$. We thus expect that the
correspondence between
normalisable modes and the poles of $D^R(\omega,k)$
in the low temperature phase is, in fact, valid for any $s=h-\bar h$.

\section{Discussion}
As pointed out in \cite{Dyson:2002nt}-\cite{Kleban2}, the energy spectrum for a system with finite entropy $S$ is
discrete, with level spacing of the order $e^{-S}$. As a consequence, the
 system will necessarily show Poincar\'{e} recurrence.
These recurrences will occur on a timescale of the order $t \sim e^{S}$.
Indeed, as we saw in section 2, for a system with discrete spectrum, the retarded propagator has
simple poles on the real axis. Therefore, the linear response
will exhibit oscillatory behaviour. In general, however, this oscillatory behaviour will be quite complicated.
Typically, one expects the correlation function for a system with finite entropy to be
a quasiperiodic function, with incommensurate frequencies.
Poincar\'{e} recurrence ensures that the evolution is unitary with no
loss of information. This is exactly the behaviour seen 
in the free fermion correlation function
(\ref{5}): at finite $L$ it is oscilating with period $L$.

Based on these remarks, it is important to understand the behaviour observed in the
strongly coupled CFT. In the strict ${\tt k}=\infty$ limit, the sole contribution to the
correlator is given by the BTZ black hole. The decay of this correlator is due to the fact that
the effective volume $L_{\rm{eff}} \simeq {\tt k}L$ is infinite, 
and thus the spectrum is continuous. 
This explains why we found complex poles corresponding to quasinormal modes of the black hole. However, for finite ${\tt k}$ the correlator must ultimately exhibit Poincar\'{e} recurrence.

In the present paper, we considered the inclusion of the contribution from thermal AdS. We found that this gives a periodic contribution, whose frequencies are the same as the normal modes
of AdS. While the addition of this contribution does prevent
the decay of the correlator at late times \cite{Maldacena:2001kr}, it is not sufficient
to produce the Poincar\'{e} recurrences at finite $\tt k$. One could consider the inclusion of the remaining members of the $SL(2, {\bf Z})$
family of solutions with torus boundary. These contributions are necessary in order
to ensure  modular invariance. However,
they will be parametrically of the same order as the thermal AdS contribution. The total correlator will still
include the decaying BTZ part with complex frequencies.
It seems that in order to see the discreteness of the energy spectrum on the CFT side, one will need to include finite ${\tt k}$ corrections
to the correlator. In this way we expect that the ${\tt k}=\infty$ (BTZ) contribution 
will  be dressed by $1/{\tt k}$ corrections, so that the correlator at finite $\tt k$  
will no longer
be  a decaying function of time. 
One can see how this may happen by recalling the case of free fermions. The correlation function
(\ref{5}) is decaying when $L=\infty$ and is periodic at finite $L$.
A somewhat similar behaviour is expected in the strongly coupled case.
As a result, the Poncar\'{e} recurrence time
will become finite and set by $L_{\rm eff}$. 
This problem certainly warrants further investigation.
In effect, there is no fundamental difference between the free and interacting case.
The free system is periodic in time, while the interacting system should exhibit Poincar\'{e} recurrences
as a quasiperiodic function. In both cases, the evolution will be unitary with no loss of information, as
expected for a system in finite volume.

We should also stress that the conformal field theory dual to
supegravity on
AdS$_3$ is very special. For a generic CFT, we do not 
expect a phase
transition even at strong coupling. It would be
interesting
therefore to find more examples of interacting CFT's in various
dimensions,
where explicit non-perturbative results can be obtained. In this
respect, the
duality between the $O(N)$ sigma model in $3$-dimensions and fields of
even
spin in AdS$_4$  might be of interest \cite{Klebanov:2002ja}. Note also
that while the explicit
computation of finite temperature Green's functions in AdS$_d$, $d>3$,
is
generally not possible, the quasinormal modes can nevertheless be
obtained numerically \cite{Horowitz:1999jd,Govindarajan:2000vq}. 
In this way, one can 
obtain quantitative, non-perturbative, information about thermal
Green's
functions also in higher dimensions. For $d>3$, the high temperature
phase
is an AdS-Schwarzschild black hole whereas thermal AdS is dominant at
low
temperature. This Hawking-Page transition is then related to linear response theory
in
the dual CFT via quasinormal  and normal modes in the two backgrounds.
\vspace*{1cm}

\noindent{\bf Acknowledgements}\\[.5ex]
We would like to thank G. Arutyunov, A. Brandhuber, B. Fine, B. Freivogel,
V. Hubeny, S. Kachru, E. Keski-Vakkuri, M. Kleban,
J. Maldacena, S. Shenker, L. Susskind, S. Theissen, E. Verlinde,
and A. Zelnikov for valuable discussions.
D.B. is grateful to the Department of Physics at Stanford University
for hospitality during the completion of this work.
This work was partially
supported by Enterprise Ireland grants IC/2001/004
and IC/2002/021. S.S. is supported by
the grant DFG-SPP 1096, Stringtheorie.

\end{document}